\documentclass[smallextended, referee]{svjour3}
\smartqed
\usepackage{amsmath}
\usepackage{cases}
\usepackage{amssymb}
\usepackage{amscd}
\usepackage{latexsym}
\usepackage{indentfirst}
\usepackage{subfigure}
\usepackage{graphicx}
\usepackage{fix-cm}
\usepackage{color}

\journalname{J. Stat. Phys.}

\begin{document}

\title{
  \bf Optimization of stochastic thermodynamic machines
}

\titlerunning{Optimization of stochastic thermodynamic machines}

\author{Yunxin Zhang}
\authorrunning{Y. Zhang}

\institute{Shanghai Key Laboratory for Contemporary Applied Mathematics, Centre for Computational Systems Biology, School of Mathematical Sciences, Fudan University, Shanghai 200433, China. Email: xyz@fudan.edu.cn.}

{ \date{\bf\large\today}}

\maketitle

\begin{abstract}
The study of stochastic thermodynamic machines is one of the main topics in nonequilibrium thermodynamics. In this study, within the framework of Fokker-Planck equation, and using the method of characteristics of partial differential equation as well as the variational method, performance of stochastic thermodynamic machines is optimized according to the external potential, with the irreversible work $W_{irr}$, or the total entropy production $\Delta S_{\rm tot}$ equivalently, reaching its lower bound. Properties of the optimal thermodynamic machines are discussed, with explicit expressions of upper bounds of work output $W$, power $P$, and energy efficiency $\eta$ are presented. To illustrate the results obtained, typical examples with optimal protocols (external potentials) are also presented.


\end{abstract}

\section{Introduction}

How we can improve the performance of thermodynamic machines is always an interesting but difficult problem in thermodynamics since the pioneer work of Carnot and Clausius \cite{Carnot1824,Clausius1856}. According to the second law of thermodynamics, the energy efficiency, defined as $\eta=W/Q_h$ with $W$ the output work and $Q_h$ the heat uptake from the hot heat bath, is not more than the Carnot efficiency $\eta_C=1-T_c/T_h$, where $T_h$ and $T_c$ are absolute temperatures of hot and cold heat baths, respectively. However, it is commonly thought the Carnot efficiency $\eta_C$ can only be achieved in quasistatic limit with output power vanished. So, in recent decades, most studies turned to discuss the efficiency at (or near) maximum power (EMP) $\eta_{\max P}$. For example, the Curzon-Ahlborn efficiency at maximum power $\eta_{CA}=1-\sqrt{T_c/T_h}$ is obtained in \cite{Yvon1955,Chambadal1957,Novikov1958,Curzon1975,Broeck2005}. The bounds of EMP, $\eta_C/2\le\eta_{\max P}\le \eta_C/(2-\eta_C)$, is obtained in \cite{Schmiedl2008Efficiency1,Esposito2010Efficiency,Izumida2011Efficiency}. By  methods of linear irreversible thermodynamics, general expressions for maximum power and maximum efficiency are obtained in \cite{Proesmans2016}. The upper bound of efficiency at arbitrary power is discussed in \cite{Holubec2016,Ryabov2016,Pietzonka2018}. For more studies which try to optimize the performance of thermodynamic machines, see \cite{Benenti2011,Izumida2011Efficiency,Golubeva2012Efficiency,Allahverdyan2013,Holubec2015,Calvo2015,Reyes2017,Polettini2017}. Or see \cite{Sekimoto2010,Seifert2012Stochastic,Mart2016Colloidal,Giuliano2017} for general reviews. Meanwhile, methods to operate heat engines infinitely close to the Carnot bound $\eta_C$ but with nonzero power are also suggested recently
\cite{Benenti2011,Allahverdyan2013,Mart2016Brownian,Lee2017Carnot,Polettini2017,Holubec2018},
though there are also studies which show that it is impossible \cite{Hondou2000Unattainability,Verley2014The}.

One of the main difficulties in the study of stochastic thermodynamic machines is that, except for few special cases, no expressions for work $W$, power $P$ and efficiency $\eta$ can be obtained explicitly \cite{Schmiedl2007,Schmiedl2008Efficiency1,Holubec2014,Holubec2015,Zhang2019}. So, in references, they are usually optimized by choosing optimal work period $t$ to get maximum power, or just through extensive numerical calculations to get optimal values of thorough protocols \cite{Then2008,Horowitz2018}. In this study, using variational method, the external time-dependent potential, as well as the boundary distributions of isothermal steps in each work cycle, will be optimized generally to achieve the optimal performance of stochastic thermodynamic machines.

The same as in \cite{Schmiedl2008Efficiency1,Holubec2014,Zhang2019}, this study considers a thermodynamic machine with potential $V(x,\tau)$, which depends on spatial (state) variable $x$ and time variable $\tau$. For overdamped cases, the time evaluation of probability $p(x,\tau)$ to find the system at position $x$ at time $\tau$ is governed by the following Fokker-Planck equation
\begin{eqnarray}\label{eq1}
\partial_\tau p(x,\tau)=-\partial_x J(x,\tau)=-\partial_x [p(x,\tau) u(x,\tau)],
\end{eqnarray}
with $J(x,\tau):=p(x,\tau) u(x,\tau)$ the flux of probability, $u(x,\tau)$ the instantaneous velocity given by $u(x,\tau):=-\partial_x [V(x,\tau)+k_BT\ln p(x,\tau)]/\xi$, and $\xi$ the friction coefficient, which satisfies Einstein relation $k_BT=\xi D$. Here $k_B$ is the Boltzmann constant, $T$ is the absolute temperature, and $D$ is the free diffusion constant.

The thermodynamic machine discussed in this study is assumed to work cyclicly, including two isothermal processes and two adiabatic transitions. In each work cycle, it performs sequently through the following four subprocesses \cite{Schmiedl2008Efficiency1,Zhang2019}. {\bf (1)} Isothermal process with high temperature $T_h$ during time interval $0<\tau<t_h$. {\bf (2)} Adiabatic transition (instantaneously) from high temperature $T_h$ to low temperature $T_c$ at time $\tau=t_h$. {\bf (3)} Isothermal process with low temperature $T_c$ during time interval $t_h<\tau<t_h+t_c$.  {\bf (4)} Adiabatic transition from low temperature $T_c$ to high temperature $T_h$ at time $\tau=t_h+t_c$. The same as in \cite{Schmiedl2008Efficiency1,Holubec2014,Zhang2019}, adiabatic transitions are idealized as sudden jumps of potential, and assumed to occur instantaneously without heat exchange. Therefore, probability $p(x,\tau)$ and system entropy do not change during adiabatic transitions. For convenience, the work period is denoted as $t:=t_h+t_c$.

It is obvious that most of physical quantities of the thermodynamic machine, including output work $W$ in each work cycle, power $P=W/t$, and energy efficiency $\eta$, are {\it functions} of duration $t_h$ (or $t_c$), and period $t=t_h+t_c$, or even {\it functions} of temperatures $T_{h/c}$, friction coefficients $\xi_{h/c}$, and diffusion constants $D_{h/c}$, and so can be optimized according to them, as have been done previously \cite{Schmiedl2008Efficiency1,Esposito2010Efficiency,Izumida2011Efficiency,Holubec2016,Golubeva2012Efficiency,Allahverdyan2013,Van2005Thermodynamic,Cleuren2009Universality,Van2012Efficiency,Wang2013Efficiency,Hooyberghs2013Efficiency}.
However, it is no doubt that $W, P$, and $\eta$ are also {\it functionals} of external potential $V(x, \tau)$ (or called protocols sometimes), probabilities $p(x,0)=p(x,t)$ and $p(x,t_h)$. One main contribution of this study is that the performance of thermodynamic machine is optimized according to potential $V(x, \tau)$, as well as to probabilities $p(x,0)$ and $p(x,t_h)$.

\section{Methods of optimization}\label{SecII}

As stated above, this study assumes that the work cycle begins from the isothermal process with high temperature $T_h$. By definition, the heat uptake from the hot heat bath during one work cycle is
\begin{eqnarray}\label{eq2}
Q_h=\int_X\left(\int_0^{t_h} \partial_\tau p(x,\tau)V(x,\tau)d\tau\right)dx.
\end{eqnarray}
In the following, we assume $X=[0,x_0]$. For other cases, such as $X=[0,+\infty)$, or $(-\infty,+\infty)$, results are similar, and corresponding examples will be presented at the end of this study. Here $x_0>0$ is a positive real number, which is used to describe the scale of the thermodynamic system, see Sec.~\ref{SubSecIIB} for some details.

From Eq.~(\ref{eq1}), and with the no flux boundary conditions at $x=0$ and $x_0$, it can be verified that potential $V(x,\tau)$ can be formally reformulated as
\begin{eqnarray}\label{eq2-1}
V(x,\tau)=\int_0^x\frac{[\xi\int_0^z\partial_\tau p(y,\tau)dy]-k_BT_h \partial_z p(z,\tau)}{p(z,\tau)}dz+C(\tau)
\end{eqnarray}
for $0<\tau<t_h$ and $0\le x\le x_0$, with $C(\tau)$ an arbitrary function of time $\tau$. Substituting $V(x,\tau)$ into Eq.~(\ref{eq2}), and through routine calculations, one can get
\begin{eqnarray}\label{eq3}
\begin{aligned}
Q_h=&
-\int_0^{x_0}\left(\int_0^{t_h} \partial_x J(x,\tau)V(x,\tau)d\tau\right)dx\cr
=&\int_0^{x_0}\left(\int_0^{t_h} J(x,\tau)\partial_x V(x,\tau)d\tau\right)dx\cr
\overset{(*)}{=}&-k_BT\int_0^{x_0}\left(\int_0^{t_h} J(x,\tau)\partial_x[\ln p(x,\tau)]d\tau\right)dx\cr
&-\int_0^{x_0}\left(\int_0^{t_h} \xi_h J(x,\tau) u(x,\tau)d\tau\right)dx\cr
=&k_BT\int_0^{x_0}\left(\int_0^{t_h} \partial_x J(x,\tau)\ln p(x,\tau)d\tau\right)dx\cr
&-\int_0^{x_0}\left(\int_0^{t_h} \xi_h p(x,\tau) u^2(x,\tau)d\tau\right)dx\cr
=:&-k_BT\int_0^{x_0}\left(\int_0^{t_h} \partial_{\tau} p(x,\tau)\ln p(x,\tau)d\tau\right)dx-W^h_{irr}\cr
\overset{(\#)}{=}&-k_BT\int_0^{x_0}\left(\int_0^{t_h} \partial_{\tau} [p(x,\tau)\ln p(x,\tau)]d\tau\right)dx-W^h_{irr}\cr
=:&T_h\Delta S-W^h_{irr},
\end{aligned}
\end{eqnarray}
where the equality $(*)$ is from the definition of instantaneous velocity $u(x,\tau):=-\partial_x [V(x,\tau)+k_BT\ln p(x,\tau)]/\xi$, and the equality $(\#)$ is because
$$\int_0^{x_0}\left(\int_0^{t_h} p(x,\tau)\partial_{\tau}[\ln p(x,\tau)] d\tau\right)dx=\int_0^{x_0}\left(\int_0^{t_h}\partial_{\tau} p(x,\tau) d\tau\right)dx=0.$$
The entropy production $\Delta S:=S(t_h)-S(0)$ with system {\it entropy} $S(\tau):=-k_B\int_0^{x_0}p(x,\tau)\ln p(x,\tau) dx$, and the {\it irreversible work} $W^h_{irr}$ is defined by
\begin{equation}\label{eq4}
\begin{aligned}
W^h_{irr}
:=&\int_0^{t_h}\int_0^{x_0}\xi_h p(x,\tau) u^2(x,\tau)dx d\tau\cr
=&\int_0^{t_h}\int_0^{x_0}\xi_h[\partial_\tau f(x,\tau)]^2/\partial_xf(x,\tau) dx d\tau.
\end{aligned}
\end{equation}
In Eq.~(\ref{eq4}), $f(x,\tau):=\int_0^xp(z,\tau)dz$ is the {\it distribution function} of the system at time $\tau$, which satisfies (see Eq.~(\ref{eq1}) and the no flux boundary condition at $x=0$)
\begin{eqnarray}\label{eq5}
\partial_\tau f(x,\tau)+u(x,\tau)\partial_x f(x,\tau)=0.
\end{eqnarray}

Eq.~(\ref{eq5}) implies that, along its {\it characteristic curves}, which are given by $dx/d\tau=u(x,\tau)$, {\it distribution function} $f(x,\tau)$ satisfies
\begin{align}
\frac{df(x,\tau)}{d\tau}&=\partial_\tau f(x,\tau)+\frac{dx}{d\tau}\cdot\partial_x f(x,\tau)\nonumber\\
&=\partial_\tau f(x,\tau)+u(x,\tau)\partial_x f(x,\tau)\nonumber\\
&=0.\nonumber
\end{align}
Which shows that $f(x,\tau)$ is constant along any {\it characteristic curve} of Eq.~(\ref{eq5}). For simplicity, this study assumes that, for any $0\le \tau\le t$, $f(x,\tau)$ is reversible as a function of $x$, or equivalently, probability $p(x,\tau)$ at any given time $\tau$ is positive at almost everywhere. Or, in other words, there is no interval $[x_1, x_2]$ with $0\le x_1<x_2\le x_0$, such that $p(x,\tau)\equiv0$ in it, {\it i.e.}, $p(x,\tau)=0$ at the most for a zero measure set \cite{Loeb2016}. But all results obtained in this study hold true for any general cases (note that, more complicated details need to be added to discuss the general cases).

Similarly, heat exchange between thermodynamic machine and cold heat bath in one work cycle can be written as $Q_c=-T_c\Delta S-W^c_{irr}$, with {\it irreversible work} $W^c_{irr}$ defined similarly. By definition, the output work during one work cycle is
\begin{equation}\label{eq6}
\begin{aligned}
W=&-\int_0^{x_0}\left(\int_0^{t} p(x,\tau)\partial_\tau V(x,\tau)d\tau\right)dx\cr
=&\int_0^{x_0}\left(\int_0^{t} \partial_\tau p(x,\tau)V(x,\tau)d\tau\right)dx\cr
=&Q_h+Q_c\cr
=&(T_h-T_c)\Delta S-(W^h_{irr}+W^c_{irr}).
\end{aligned}
\end{equation}

Two strategies to improve work $W$ will be presented in this study. {\bf (I)} For given probabilities $p_0(x):=p(x,0)$ and $p_1(x):=p(x,t_h)$, {\it i.e.} for given $\Delta S$, find the minimum value of {\it irreversible work} $W^{h/c}_{irr}$ by optimization according to potential $V(x,\tau)$. {\bf (II)} For given probability $p_0(x)$, find the maximum value of $\Delta S$ by optimization according to probability $p_1(x)$.

\subsection{Optimization of distribution function $f(x,\tau)$}
From Eq.~(\ref{eq4}), it can be shown that the variation of {\it irreversible work} $W^h_{irr}$ according to the {\it distribution function} $f(x,\tau)$ is as follows,
\begin{small}
\begin{eqnarray}\label{eq7}
\delta W^h_{irr}=\int_0^{t_h}\int_0^{x_0}2\xi_h\left[\frac{\partial_{\tau}f}{\partial_xf}
\partial_{x}\left(\frac{\partial_{\tau}f}{\partial_xf}\right)
-\partial_{\tau}\left(\frac{\partial_{\tau}f}{\partial_xf}\right)\right]
\delta f\,dx d\tau,
\end{eqnarray}
\end{small}
where $\delta f$ is an arbitrary variation of $f(x,\tau)$, which satisfies
$$\delta f(0, \tau)=\delta f(x_0,\tau)=\delta f(x,0)=\delta f(x,t_h)=0.$$
The reason of these zero boundary conditions is that $f(x,\tau)$ satisfies $f(0,\tau)\equiv0, f(x_0,\tau)\equiv1$, and $f(x,0)=\int_0^xp_0(z)dz, f(x,t_h)=\int_0^xp_1(z)dz$.

Since $\delta W^h_{irr}=0$ at the minimum of $W^h_{irr}$ (as a functional of {\it distribution function} $f(x,\tau)$) for any variation $\delta f$, Eq.~(\ref{eq7}) indicates that, for the optimal {\it distribution function} $f(x,\tau)=f^*(x,\tau)$,
$$
\left(\frac{\partial_{\tau}f^*}{\partial_{x}f^*}\right)
\partial_{x}\left(\frac{\partial_{\tau}f^*}{\partial_{x}f^*}\right)
-\partial_{\tau}\left(\frac{\partial_{\tau}f^*}{\partial_{x}f^*}\right)=0.
$$
Due to Eq.~(\ref{eq5}), $\partial_{\tau}f/\partial_{x}f=-u$. Therefore,
\begin{eqnarray}\label{eq8}
\partial_{\tau}u^*(x,\tau)+u^*(x,\tau)\partial_{x}u^*(x,\tau)=0.
\end{eqnarray}

Eq.~(\ref{eq8}) shows that, for the optimal {\it distribution function} $f^*(x,\tau)$, the slope $u^*(x,\tau)$ of its {\it characteristic curves} satisfies
$$
\frac{du^*(x,\tau)}{d\tau}
=\partial_{\tau}u^*(x,\tau)+\frac{dx}{d\tau}\partial_{x}u^*(x,\tau)=0.
$$
It means that, the slope $u^*(x,\tau)$ of any {\it characteristic curves} of the optimal {\it distribution function} $f^*(x,\tau)$ is constant. Therefore, {\it characteristic curves} of $f^*(x,\tau)$, which are given by $dx/d\tau=u(x,\tau)$, are all straight lines.

For convenience, we denote $f_0(x):=f(x,0)$, $f_1(x):=f(x,t_h)$, and define a map $\Gamma: x\in [0, x_0]\rightarrow \Gamma(x)\in [0, x_0]$ which satisfies $f_0(x)=f_1(\Gamma(x))$, or equivalently
$$\Gamma(x)=f_1^{-1}(f_0(x)).$$
Note, for extreme cases that at least one of functions $f_0(x)$ and $f_1(x)$ is irreversible, map $\Gamma(x)$ can also be well defined but with some complicated and tedious details, so will not be presented here. Obviously, map $\Gamma$ satisfies $\Gamma(0)=0, \Gamma(x_0)=x_0$, and it increases monotonically with $x$, since $\Gamma'(x)=p_0(x)/p_1(\Gamma(x))>0$.

The above analysis shows that, for optimal {\it distribution function} $f^*(x,\tau)$, the {\it characteristic curve} of Eq.~(\ref{eq5}), which is initiated from any $z\in [0,x_0]$ at time $\tau=0$, satisfies
\begin{eqnarray}\label{eq9}
x(z,\tau)=z+u^*(x,\tau)\tau=z+[\Gamma(z)-z]\tau/t_h.
\end{eqnarray}
Which means that the {\it characteristic curve} of Eq.~(\ref{eq5}), which begins from any $z\in [0,x_0]$ at time $\tau=0$, is a straight line connecting point $(z,0)$ and point $(\Gamma(z),t_h)$ in the 2-dimensional $x$-$\tau$ plane.
For any given $z\in [0,x_0]$ at time $\tau=0$, the position $x(z,\tau)$ obtained by Eq.~(\ref{eq9}) satisfies $f^*(x(z,\tau),\tau)=f_0(z)\equiv f_1(\Gamma(z))$, see Eq.~(\ref{eq5}). For convenience, we denote the inverse function of $x(z,\tau)$ by $z(x,\tau)$. Then, for any given position $x\in [0,x_0]$ and time $0\le\tau\le t_h$, $z(x,\tau)$ satisfies $f^*(x,\tau)=f_0(z(x,\tau))\equiv f_1(\Gamma(z(x,\tau)))$. Finally, Eq.~(\ref{eq9}) is actually obtained by definition from the following ordinary differential equation,
$dx(\tau)/d\tau=u^*(x,\tau)$, with $x(0)=z$. While $u^*(x,\tau)\equiv[\Gamma(z)-z]/t_h$ for the optimal {\it distribution function} $f^*(x,\tau)$.

Using function $x(z,\tau)$ and its inverse function $z(x,\tau)$, probability $p^*(x,\tau)$ corresponding to the optimal {\it distribution function} $f^*(x,\tau)$ can be obtained by $p^*(x,\tau)=\partial_x f^*(x,\tau)=\partial_x f_0(z(x,\tau))=p_0(z(x,\tau))\partial_x z(x,\tau)$.
From Eq.~(\ref{eq4}), it can be shown that, the {\it irreversible work} $W^h_{irr}$ corresponding to the optimal {\it distribution function} $f^*(x,\tau)$ is {\small
\begin{equation}\label{eq11}
\begin{aligned}
W^{h*}_{irr}=&\int_0^{t_h}\int_0^{x_0}\xi_h p^*(x,\tau) [u^*(x,\tau)]^2dx d\tau\cr
=&\int_0^{t_h}\int_0^{x_0}\xi_h p_0(z(x,\tau))\partial_x z(x,\tau)\left[\frac{\Gamma(z(x,\tau))-z(x,\tau)}{t_h}\right]^2 dx d\tau\cr
=&\int_0^{t_h}\int_0^{x_0}\xi_h p_0(z)\left[\frac{\Gamma(z)-z}{t_h}\right]^2 dz d\tau\cr
=&\frac{\xi_h}{t_h}\int_0^{x_0}p_0(z)[\Gamma(z)-z]^2 dz.
\end{aligned}
\end{equation}}
Here, the third equality is obtained by change of variable $z=z(x,\tau)$.

Due to $f_0(y)=f_1(\Gamma(y))$, $p_0(y)=p_1(\Gamma(y))\Gamma'(y)$ or $p_1(y)=p_0(\Gamma^{-1}(y))/\Gamma'(\Gamma^{-1}(y))$. So, it can be shown that
\begin{equation}\label{eq12}
\begin{aligned}
W^{c*}_{irr}=&\frac{\xi_c}{t_c}\int_0^{x_0}p_1(y)[y-\Gamma^{-1}(y)]^2 dy\cr
=&\frac{\xi_c}{t_c}\int_0^{x_0}\frac{p_0(\Gamma^{-1}(y))}{\Gamma'(\Gamma^{-1}(y))}[y-\Gamma^{-1}(y)]^2 dy\cr
=&\frac{\xi_c}{t_c}\int_0^{x_0}p_0(z)[\Gamma(z)-z]^2 dz,
\end{aligned}
\end{equation}
Here, the last equality is obtained by change of variable $y=\Gamma(z)$.
Meanwhile, it can be verified that
$$
\begin{aligned}
S(t_h)=&-k_B\int_0^{x_0}p_1(y)\ln p_1(y)dy\cr
=&-k_B\int_0^{x_0}\frac{p_0(\Gamma^{-1}(y))}{\Gamma'(\Gamma^{-1}(y))}
\ln\left(\frac{p_0(\Gamma^{-1}(y))}{\Gamma'(\Gamma^{-1}(y))}\right)dy\cr
=&-k_B\int_0^{x_0}\frac{p_0(z)}{\Gamma'(z)}
\ln\left(\frac{p_0(z)}{\Gamma'(z)}\right)\Gamma'(z)dz\cr
=&-k_B\int_0^{x_0}p_0(z)\left(\ln p_0(z)-\ln \Gamma'(z)\right)dz\cr
=&S(0)+k_B\int_0^{x_0}p_0(z)\ln \Gamma'(z)dz.
\end{aligned}
$$
So, by definition,
\begin{equation}\label{eq14}
\Delta S=S(t_h)-S(0)=k_B\int_0^{x_0}p_0(x)\ln \Gamma'(x) dx.
\end{equation}
From Eqs.~(\ref{eq3},\ref{eq11},\ref{eq14}), with the optimal {\it distribution function} $f^*(x,\tau)$,
\begin{equation}\label{eq15}
\begin{aligned}
Q_h^*=k_BT_h\int_0^{x_0}p_0(x)\ln \Gamma'(x) dx-\frac{\xi_h}{t_h}\int_0^{x_0}p_0(x)[\Gamma(x)-x]^2 dx,
\end{aligned}
\end{equation}
and from Eqs.~(\ref{eq6},\ref{eq11},\ref{eq12},\ref{eq14}),{\small
\begin{equation}\label{eq16}
\begin{aligned}
W^*=&k_B(T_h-T_c)\int_0^{x_0}p_0(x)\ln \Gamma'(x) dx-\left(\frac{\xi_h}{t_h}+\frac{\xi_c}{t_c}\right)\int_0^{x_0}p_0(x)[\Gamma(x)-x]^2 dx\cr
=&k_B(T_h-T_c)\int_0^{x_0}p_0(x)\ln \Gamma'(x) dx-\frac{1}{t}\left(\frac{\xi_h}{\bar{t}_h}+\frac{\xi_c}{\bar{t}_c}\right)\int_0^{x_0}p_0(x)[\Gamma(x)-x]^2 dx,
\end{aligned}
\end{equation}}
where $\bar{t}_h:=t_h/t, \bar{t}_c:=t_c/t$. Note, $\bar{t}_h+\bar{t}_c\equiv1$.

Generally, in one work cycle, the total entropy production of the system can be obtained by $\Delta S_{\rm tot}=-\int_0^t[\int_X\partial_\tau p(x,\tau)V(x,\tau)dx]/T(\tau)d\tau$ \cite{Brandner2015,Proesmans2015}.
Within the framework of Fokker-Planck equation, $\Delta S_{\rm tot}$ can be reformulated as $\Delta S_{\rm tot}=\int_0^t[\int_X\xi p(x,\tau)u^{2}(x,\tau)dx]/T(\tau)d\tau$.
For the particular case with optimal {\it distribution function} $f^*(x,\tau)$, it can be shown that the lower bound of total entropy production is
\begin{eqnarray*}
\Delta S_{\rm tot}^*&=&-\frac{Q_h^*}{T_h}-\frac{Q_c^*}{T_c}\cr
&=&\left(\frac{\xi_h}{t_hT_h}+\frac{\xi_c}{t_cT_c}\right)\int_0^{x_0}p_0(x)[\Gamma(x)-x]^2 dx\cr
&=&k_B\left(\frac{1}{t_hD_h}+\frac{1}{t_cD_c}\right)\int_0^{x_0}p_0(x)[\Gamma(x)-x]^2 dx.
\end{eqnarray*}

One can easily show that, with optimal {\it distribution function} $f^*(x,\tau)$,
\begin{equation}\label{eq16-1}
Q_h^*=T_h\left(\Delta S-\frac{\Sigma_h}{t_h}\right),\quad
Q_c^*=T_c\left(-\Delta S-\frac{\Sigma_c}{t_c}\right),
\end{equation}
where $\Sigma_{h,c}=\left.k_B\int_0^{x_0}p_0(x)[\Gamma(x)-x]^2 dx\right/D_{h,c}$. Eq. (\ref{eq16-1}) looks the same as the one assumed in discussions of the low dissipation cases \cite{Esposito2010Efficiency,Izumida2011Efficiency,Calvo2015,Gonzalezayala2016,Reyes2017}. However, Eq. (\ref{eq16-1}) holds for any durations $t_{h}$ and $t_c$, and $\Sigma_{h,c}$ may be very large, though they are corresponding to the least values of entropy production. While in discussions of low dissipation cases, expressions like Eq. (\ref{eq16-1}) are only the first order of their real values, and are reasonable for larger durations $t_h$ and $t_c$.

\subsection{Properties of thermodynamic machine with optimal distribution function $f^*(x,\tau)$}\label{SubSecIIB}

To discuss the influence of the scale of $x_0$, normalized probabilities $\bar{p}_{0/1}(x)$,  normalized {\it distribution function}s $\bar{f}_{0/1}(x)$, and the corresponding map $\bar{\Gamma}(x)$ are defined as follows,
\begin{equation}\label{eq17}
\begin{aligned}
&\bar{p}_{0/1}(x):= x_0p_{0/1}(x_0x), \ \bar{f}_{0/1}(x):=f_{0/1}(x_0x), \ \bar{\Gamma}(x):=\Gamma(x_0x)/x_0,
\end{aligned}
\end{equation}
with $0\le x\le1$. It can be easily verified that $\int_0^1\bar{p}_{0/1}(x)=1$, $\bar{\Gamma}(0)=0$, $\bar{\Gamma}(1)=1$, $\bar{f}_{0/1}(0)=0$, $\bar{f}_{0/1}(1)=1$, and $\bar{f}'_{0/1}(x)=\bar{p}_{0/1}(x)$, $\bar{f}_{0}(x)=\bar{f}_{1}(\bar{\Gamma}(x))$.

From Eqs.~(\ref{eq15},\ref{eq17}),
\begin{eqnarray}\label{eq18}
Q_h^*&=&k_BT_h\int_0^{1}x_0p_0(x_0z)\ln \Gamma'(x_0z) dz-\frac{\xi_h}{t_h}\int_0^{1}x_0p_0(x_0z)[\Gamma(x_0z)-x_0z]^2 dz,\cr
&=&k_BT_h\int_0^1\bar{p}_0(z)\ln \bar{\Gamma}'(z) dz-\frac{\xi_h x_0^2}{t_h}\int_0^1\bar{p}_0(z)[\bar{\Gamma}(z)-z]^2 dz.
\end{eqnarray}
Similarly, from Eqs.~(\ref{eq16},\ref{eq17}),{\small
\begin{equation}\label{eq19}
\begin{aligned}
W^*=&k_B(T_h-T_c)\int_0^1\bar{p}_0(x)\ln \bar{\Gamma}'(x) dx-\frac{x_0^2}{t}\left(\frac{\xi_h}{\bar{t}_h}+\frac{\xi_c}{\bar{t}_c}\right)\int_0^1\bar{p}_0(x)[\bar{\Gamma}(x)-x]^2 dx.
\end{aligned}
\end{equation}}
Note that $\bar{t}_c=1-\bar{t}_h$. Eq.~(\ref{eq19}) indicates that the value of work $W^*$ depends on parameters $t, \bar{t}_h$ (or $\bar{t}_c$), as well as functions $\bar{\Gamma}(x)$ and $\bar{p}_0(x)$, {\it i.e.} $W^*=W^*(\bar{t}_{h/c}, t; \bar{\Gamma}(x), \bar{p}_0(x))$.

From Eq.~(\ref{eq19}), the {\it stalling time} of the machine, defined by $W^*(t^{\rm st})=0$ (see \cite{Schmiedl2008Efficiency1}), is as follows,
\begin{eqnarray}\label{eq19-2}
t^{\rm st}=\frac{\left(\frac{\xi_h}{\bar{t}_h}+\frac{\xi_c}{\bar{t}_c}\right)x_0^2\int_0^1\bar{p}_0(x)[\bar{\Gamma}(x)-x]^2 dx}{k_B(T_h-T_c)\int_0^1\bar{p}_0(x)\ln \bar{\Gamma}'(x) dx}.
\end{eqnarray}
For given functions $\bar{\Gamma}(x)$, $\bar{p}_0(x)$, and period $t$, the work $W^*$ reaches its maximum when $\bar{t}_{h/c}=\sqrt{\xi_{h/c}}/(\sqrt{\xi_h}+\sqrt{\xi_c})$. This can be easily obtained by minimizing ${\xi_h}/{\bar{t}_h}+{\xi_c}/{\bar{t}_c}$ under constraint $\bar{t}_h+\bar{t}_c=1$. It can be shown that  $\max_{0<\bar{t}_h,\bar{t}_c<1; \bar{t}_h+\bar{t}_c=1}W^*=$
\begin{equation}\label{19-1}
\begin{aligned}
k_B(T_h-T_c)\int_0^1\bar{p}_0(x)\ln \bar{\Gamma}'(x) dx-\frac{(\sqrt{\xi_h}+\sqrt{\xi_c})^2x_0^2}{t}\int_0^1\bar{p}_0(x)[\bar{\Gamma}(x)-x]^2 dx.
\end{aligned}
\end{equation}

It can also be shown that power $P(t):=W^*/t$ reaches its maximum when period $t=t^*:=2t^{\rm st}$ (see Appendix). The corresponding work for period $t=t^*$ is
\begin{eqnarray}\label{eq22}
W^*_{\max P}=\frac{k_B(T_h-T_c)}{2}\int_0^1\bar{p}_0(x)\ln \bar{\Gamma}'(x) dx,
\end{eqnarray}
and the heat absorbed from hot heat bath is
\begin{equation}\label{eq22-1}
\begin{aligned}
Q^*_{h,\max P}=&k_BT_h\int_0^1\bar{p}_0(x)\ln \bar{\Gamma}'(x) dx-\frac{x_0^2\xi_h}{t_h}\int_0^1\bar{p}_0(x)[\bar{\Gamma}(x)-x]^2 dx\cr
=&k_BT_h\int_0^1\bar{p}_0(x)\ln \bar{\Gamma}'(x) dx\cr
&-\frac{(\xi_h/t_h)(\xi_h/t_h+\xi_c/t_c)x_0^2}{\xi_h/t_h+\xi_c/t_c}
\int_0^1\bar{p}_0(x)[\bar{\Gamma}(x)-x]^2 dx\cr
&=k_B\int_0^1\bar{p}_0(x)\ln \bar{\Gamma}'(x) dx\left(T_h
-\frac{(T_h-T_c)\xi_h/t_h}{2(\xi_h/t_h+\xi_c/t_c)}\right).
\end{aligned}
\end{equation}
So the efficiency at maximum power is
\begin{eqnarray}\label{eq23}
\eta^*_{\max P}=\frac{W^*_{\max P}}{Q^*_{h,\max P}}
=\frac{(T_h-T_c)/2}{T_h
-\frac{(T_h-T_c)\xi_h/t_h}{2(\xi_h/t_h+\xi_c/t_c)}}
=\frac{\eta_C}{2-\frac{\eta_C\xi_h/t_h}{\xi_h/t_h+\xi_c/t_c}}.
\end{eqnarray}

As has been found previously, $\eta_C/2<\eta^*_{\max P}<\eta_C/(2-\eta_C)$ \cite{Schmiedl2008Efficiency1,Esposito2010Efficiency,Izumida2011Efficiency,Johal2017Heat,Giuliano2017,Reyes2017}. It tends to lower bound $\eta_C/2$ if $(t_c\xi_h)/(t_h\xi_c)\to0$, and tends to upper bound $\eta_C/(2-\eta_C)$ if $(t_h\xi_c)/(t_c\xi_h)\to0$. For particular cases with $\bar{t}_{h/c}=\sqrt{\xi_{h/c}}/(\sqrt{\xi_h}+\sqrt{\xi_c})$, with which  work $W^*$ is maximized ($t^*$ and $t^{\rm st}$ are minimized), $\eta^*_{\max P}=\eta_C/[2-\eta_C\sqrt{\xi_{h}}/(\sqrt{\xi_h}+\sqrt{\xi_c})]$. If $\xi_h=\xi_c$, $\eta^*_{\max P}=\eta_C/[2-\eta_C t_c/(t_c+t_h)]$. It can be easily known that, for $t_h=t_c=t/2$, the optimal probability $p^*(x,\tau)$ is mirror symmetrical to time $t/2$, {\it i.e.}, $p^*(x, t/2-\tau)=p^*(x, t/2+\tau)$ for $0\le\tau\le t/2$. For such cases, $\eta^*_{\max P}=\eta_C/[2-\eta_C \xi_h/(\xi_h+\xi_h)]$.

It can be verified that $\eta^*_{\max P}\le\eta_{CA}$ if and only if $[(\xi_c/t_c)/(\xi_h/t_h)]^2\ge T_c/T_h$.
This condition is equivalent to the one given in \cite{Schmiedl2008Efficiency1}, where the Curzon-Ahlborn efficiency $\eta_{CA}$ is recovered if $\alpha=\alpha_{CA}\equiv1/(1+\sqrt{T_c/T_h})$, with parameter $\alpha=(\xi_h/t_h)/(\xi_h/t_h+\xi_c/t_c)$ in this study, see the above Eq.~(\ref{eq23}) and the Eq.~(31) in \cite{Schmiedl2008Efficiency1}. For  special cases with $\bar{t}_{h}=\bar{t}_{c}=1/2$, inequality $[(\xi_c/t_c)/(\xi_h/t_h)]^2\ge T_c/T_h$ reduces to $(\xi_c/\xi_h)^2\ge T_c/T_h$. While for cases with $\bar{t}_{h/c}=\sqrt{\xi_{h/c}}/(\sqrt{\xi_h}+\sqrt{\xi_c})$, it reduces to $\xi_c/\xi_h\ge T_c/T_h$. So, with optimal {\it distribution function} $f^*(x,\tau)$ and optimal ratio of duration $t_h/t_c=\sqrt{\xi_h/\xi_h}$, and if diffusion constant $D_h=D_c$, which means $T_c/T_h=\xi_c/\xi_h$, then the efficiency at maximum power $\eta^*_{\max P}$ is exactly the Curzon-Ahlborn efficiency $\eta_{CA}$.

With optimal {\it distribution function} $f^*(x,\tau)$, the efficiency for any cyclic period $t>t^{\rm st}$ is, see Eqs.~(\ref{eq18},\ref{eq19}),
\begin{eqnarray}\label{eq24}
\eta^*=\frac{W^*}{Q_h^*}=1-\frac{T_c}{T_h}\cdot
\frac{t+2\Omega(\bar{p}_0,\bar{\Gamma})\hat{t}_c/\bar{t}_c}{t-2\Omega(\bar{p}_0,\bar{\Gamma})\hat{t}_h/\bar{t}_h},
\end{eqnarray}
where $\hat{t}_{h/c}:= x_0^2/(2D_{h/c})=\xi_{h/c} x_0^2/(2k_BT_{h/c})$ is the {\it characteristic time} of free diffusion, $\Omega(\bar{p}_0,\bar{\Gamma})$ is a functional of $\bar{p}_0(x)$ and $\bar{\Gamma}(x)$,
{\small
$$\Omega(\bar{p}_0,\bar{\Gamma})=\left.\left(\int_0^1\bar{p}_0(x)[\bar{\Gamma}(x)-x]^2
dx\right)\right/\left(\int_0^1\bar{p}_0(x)\ln \bar{\Gamma}'(x) dx\right).
$$ }
Eq.~(\ref{eq24}) shows that efficiency $\eta^*$ increases with both durations (fractions) $t_h (\bar{t}_h)$ and $t_c (\bar{t}_c)$, while decreases with both {\it characteristic times} $\hat{t}_h$ and $\hat{t}_c$. Meanwhile, $\eta^*$ increases with cyclic period $t$ but decreases with scale parameter $x_0$. Moreover, it can be shown from Eq.~(\ref{eq24}) that
$$
\eta^*\le \eta_C-2(1-\eta_C)\left(\frac{\hat{t}_h}{\bar{t}_h}+\frac{\hat{t}_h}{\bar{t}_h}\right)\frac{\Omega(\bar{p}_0,\bar{\Gamma})}{t},
$$
where the equality holds if $\Omega(\bar{p}_0,\bar{\Gamma})=0$, or $t\to\infty$.

Denote $P^*:=W^*(t^*)/t^*\equiv W^*_{\max P}/t^*$ as the maximum of power $P(t):=W^*(t)/t$, then for any $t\ge t^{\rm st}$,
\begin{eqnarray}\label{eq24-1}
\frac{P(t)}{P^*}=\frac{t^*}{t}\left(2-\frac{t^*}{t}\right)
=1-\left(\frac{t^*}{t}-1\right)^2,
\end{eqnarray}
which equals to zero when $t=t^{\rm st}=t^*/2$ or $t\to\infty$, and reaches it maximum 1 when $t=t^*$, see Eq.~(\ref{S2}) in Appendix. Meanwhile,  efficiency $\eta^*$ can be reformulated as (see Eq.~(\ref{S3}) in Appendix)
\begin{eqnarray}\label{eq24-2}
\eta^*=\left.\eta_C\left(2-\frac{t^*}{t}\right)\right/\left(2-\frac{\eta_C{\xi_h}/{\bar{t}_h}}{{\xi_h}/{\bar{t}_h}+{\xi_c}/{\bar{t}_c}}\cdot\frac{t^*}{t}\right).
\end{eqnarray}
It can be easily shown that $d\eta^*/dt>0$, so efficiency $\eta^*$ increases monotonically from 0 to $\eta_C$ as the cyclic period $t$ increases from {\it stalling time} $t^{\rm st}=t^*/2$ to infinity. For $t=t^*$, {\it i.e.}, when power $P(t)$ reaches its maximum $P^*$, efficiency $\eta^*$ get its value $\eta_{\max P}^*$ as given in Eq.~(\ref{eq23}).

From Eqs.~(\ref{eq24-1}, \ref{eq24-2}), the ratio ${P(t)}/{P^*}$ can be reformulated as a function of efficiency $\eta^*$,
\begin{eqnarray}\label{eq24-3}
\frac{P(t)}{P^*}
=\frac{4\left(\frac{1}{\eta_C}-\frac{{\xi_h}/{\bar{t}_h}}{{\xi_h}/{\bar{t}_h}+{\xi_c}/{\bar{t}_c}}\right)\left(1-\frac{\eta^*}{\eta_C}\right)\eta^*}{\left(1-\frac{{\xi_h}/{\bar{t}_h}}{{\xi_h}/{\bar{t}_h}+{\xi_c}/{\bar{t}_c}}\eta^*\right)^2},
\end{eqnarray}
which equals to zero if $\eta^*=0$ or $\eta^*=\eta_C$, and reaches its maximum 1 when $\eta^*=\eta_{\max P}^*$ as given in Eq.~(\ref{eq23}).
It can be shown that ${P(t)}/{P^*}$ increases with $\eta^*$ for $0\le\eta^*\le \eta_{\max P}^*$, while decreases with $\eta^*$ for $\eta_{\max P}^*\le\eta^*\le \eta_C$.

It can also be shown that (see Appendix),
\begin{eqnarray}\label{eq24-4}
\frac{W^*(t)}{W^*_{\max P}}
=2-\frac{t^*}{t}=\frac{2\left(1-\frac{{\xi_h}/{\bar{t}_h}}{{\xi_h}/{\bar{t}_h}+{\xi_c}/{\bar{t}_c}}\eta_C\right)\eta^*}{\left(1-\frac{{\xi_h}/{\bar{t}_h}}{{\xi_h}/{\bar{t}_h}+{\xi_c}/{\bar{t}_c}}\eta^*\right)\eta_C}.
\end{eqnarray}
Therefore, $W^*(t)/W^*_{\max P}$ increases from 0 to 2 as period $t$ increases from {\it stalling time } $t^{\rm st}=t^*/2$ to infinity, or equivalently as efficiency $\eta^*$ increases from 0 to $\eta_C$. $W^*(t)/W^*_{\max P}=1$, {\it i.e.,} $W^*(t)=W^*_{\max P}$, if and only if $t=t^*$, or $\eta^*=\eta_{\max P}^*$ equivalently. As implied by Eq.~(\ref{eq22}), $W^*_{\max P}=W^*(t\to\infty)/2$.

Finally, from Eqs.~(\ref{eq24-1}, \ref{eq24-4}), it can be obtained that
\begin{eqnarray}\label{eq24-5}
\frac{P(t)}{P^*}
=\frac{W^*(t)}{W^*_{\max P}}\left(2-\frac{W^*(t)}{W^*_{\max P}}\right).
\end{eqnarray}
So, ratio $P(t)/P^*$ increases first and then decreases with ratio $W^*(t)/W^*_{\max P}$.

\subsection{Optimal probability density $p_1(x)$}
From Eq.~(\ref{eq22}), the variation of $W^*_{\max P}$ according to map $\bar{\Gamma}(x)$ is,
\begin{eqnarray}\label{eq25}
\delta W^*_{\max P}=-\frac{k_B(T_h-T_c)}{2}\int_0^1\partial_x\left(\frac{\bar{p}_0(x)}{\bar{\Gamma}'(x)}\right) \delta\bar{\Gamma}(x) dx,
\end{eqnarray}
where $\delta\bar{\Gamma}(x)$ is a (small) arbitrary variation of function $\bar{\Gamma}(x)$. Due to boundary conditions $\bar{\Gamma}(0)=0$ and $\bar{\Gamma}(1)=1$, variation $\delta\bar{\Gamma}(x)$ satisfies $\delta\bar{\Gamma}(0)=\delta\bar{\Gamma}(1)=0$.

Eq.~(\ref{eq25}) indicates that $W^*_{\max P}$ reaches its maximum when $\bar{\Gamma}(x)=\int_0^x \bar{p}_0(z) dz$, or equivalently when $\Gamma(x)=x_0\int_0^x p_0(z) dz$, see Eq.~(\ref{eq17}). Notice that, $\bar{\Gamma}(x)=\int_0^x \bar{p}_0(z) dz$ means $\bar{p}_1(x)\equiv1$ and $\bar{f}_1(x)=x$, while $\Gamma(x)=x_0\int_0^x p_0(z) dz$ means $p^*_1(x)\equiv1/x_0$ and $f^*_1(x)\equiv x/x_0$. In any way,
\begin{eqnarray}\label{eq26}
\max_{\bar{\Gamma}(x)}W^*_{\max P}=
\frac{k_B(T_h-T_c)}{2}\int_0^1\bar{p}_0(x)\ln \bar{p}_0(x) dx.
\end{eqnarray}

\subsection{Explanatory notes}
As mentioned before, work $W$ is not only a {\it function} of period $t$ and duration $t_h$ (or $t_c$), but also a {\it functional} of the potential $V(x,\tau)$ and probability $p_0(x)$ (or $p_1(x)$ equivalently). So there are many ways to do optimization. Optimization according to {\it distribution function} $f(x,\tau)$ is actually equivalent to optimization according to potential $V(x,\tau)$,  and optimization according to map $\bar{\Gamma}(x)$ is equivalent to optimization according to probability $p_1(x)$. In references, ratio $t_h/t_c$ is usually optimized to get the maximum of work $W$, while period $t$ is usually optimized to get the maximum of power $P$. However, in this study, for any given work period $t$, the potential $V(x,\tau)$ is optimized to get the minimum of entropy production $\Delta S_{\rm tot}$, or equivalently to get the maximum of work $W$.

From Eq.~(\ref{eq17}), $\int_0^1\bar{p}_0(x)\ln \bar{p}_0(x) dx=\ln x_0+\int_0^{x_0}p_0(x)\ln p_0(x) dx$. Therefore, $\max_{\bar{\Gamma}(x)}W^*_{\max P}$ increases (logarithmically) with scale parameter $x_0$ if the initial entropy $-k_B\int_0^{x_0}p_0(x)\ln p_0(x) dx$ is fixed, see Eq.~(\ref{eq26}). However, to keep power $P$ to be maximum, the  {\it optimal period} $t^*=2t^{\rm st}$ should increase like $x_0^2$, see Eq.~(\ref{eq19-2}). Therefore, the corresponding value of maximum power $[\max_{\bar{\Gamma}(x)}W^*_{\max P}]/t^*$  actually decreases with scale parameter $x_0$.

For convenience, we denote the spatial coordinate of a point on the {\it characteristic curve} of Eq.~(\ref{eq5}),  which begins from $z$ at time $\tau=0$, by $\Gamma(z,\tau)$, or $\Gamma_{\tau}(z)$ for simplicity. That is to say, $\Gamma(z,\tau)\equiv\Gamma_{\tau}(z):=x(z,\tau)$  with $x(z,\tau)$ satisfies Eq.~(\ref{eq9}). Particularly, by definition $\Gamma(x)=\Gamma(x,t_h)$. In the following, methods to get the optimal probability $p^*(x,\tau)$ and optimal potential $V^*(x,\tau)$, which correspond to the optimal {\it distribution function} $f^*(x,\tau)$, will be presented for only $0\le\tau\le t_h$. For $t_h\le\tau\le t=t_h+t_c$, methods are almost the same.

By definition, $df(x,\tau)/d\tau=0$ along any {\it characteristic curves} of Eq.~(\ref{eq5}). This means that {\it distribution function} $f(x,\tau)$ satisfies $f_0(x)\equiv f(x,0)=f(\Gamma(x,\tau),\tau)=f(\Gamma(x,t_h),t_h)\equiv f_1(\Gamma(x,t_h))$ for any $0\le\tau\le t_h$. Therefore, $f(x,\tau)=f_0(\Gamma_{\tau}^{-1}(x))=\int_{0}^{\Gamma_{\tau}^{-1}(x)}p_0(z)d\,z$.
By taking derivatives on both side of $f_0(x)=f(\Gamma(x,\tau),\tau)$, one gets $p_0(x)=p(\Gamma(x,\tau),\tau)\partial_x\Gamma_{\tau}(x)$. So
\begin{eqnarray}\label{eq28}
p(x,\tau)=p_0(\Gamma_{\tau}^{-1}(x))/\left.\partial_z\Gamma_{\tau}(z)\right|_{z=\Gamma_{\tau}^{-1}(x)}.
\end{eqnarray}
Particularly, for $\tau=t_h$,  $p_1(x)=p_0(\Gamma^{-1}(x))/\Gamma'(\Gamma^{-1}(x))$.

By definition $-\partial_x[V(x,\tau)+k_BT \ln p(x,\tau)]/\xi_h=u(x,\tau)$, so
$V(x,\tau)=-\xi_h\int_0^x u(z,\tau) d\,z-k_BT \ln p(x,\tau)+C(\tau)$, with $C(\tau)$ an arbitrary function of time $\tau$. For cases with optimal {\it distribution function} $f^*(x,\tau)$, the instantaneous velocity    $u^*(z,\tau)=[\Gamma(\Gamma_{\tau}^{-1}(z))-\Gamma_{\tau}^{-1}(z)]/t_h$. Therefore, the optimal potential is {\small
\begin{equation}\label{eq29}
\begin{aligned}
V^*(x,\tau)=&-\xi_h\int_0^x u^*(z,\tau) d\,z-k_BT \ln p^*(x,\tau)+C(\tau)\cr
=&-\xi_h\int_0^x \frac{\Gamma(\Gamma_{\tau}^{-1}(z))-\Gamma_{\tau}^{-1}(z)}{t_h} d\,z-k_BT \ln p^*(x,\tau)+C(\tau)\cr
=&-\xi_h\int_0^{\Gamma_{\tau}^{-1}(x)} \frac{\Gamma(y)-y}{t_h}\partial_y\Gamma_{\tau}(y) d\,y-k_BT \ln p^*(x,\tau)+C(\tau)\cr
=&-\xi_h\int_0^{\Gamma_{\tau}^{-1}(x)} [\Gamma(y)-y]\left[1+\frac{(\Gamma'(y)-1)\tau}{t_h}\right] d\,y
-k_BT\ln p_0(\Gamma_{\tau}^{-1}(x))\cr
&+k_BT\ln \left(1+\frac{[\Gamma'(\Gamma_{\tau}^{-1}(x))-1]\tau}{t_h}\right)+C(\tau).
\end{aligned}
\end{equation}}
Here the last equality is obtained from Eqs.~(\ref{eq9}, \ref{eq28}) and $\Gamma(z,\tau)=x(z,\tau)$.

Since thermodynamic machines work cyclically, function $C(\tau)$ should satisfies $C(\tau+t)=C(\tau)$ for any time $\tau$. One can easily find that {\it irreversible works} $W_{irr}^{h/c*}$ are independent of the choice of function $C(\tau)$, so in all the following illustrative examples, we always set $C(\tau)\equiv0$ simply.

\section{Illustrative examples with optimal control}
To illustrate the results obtained in this study, we present several examples of thermodynamic machine with the optimal {\it distribution function} $f^*(x,\tau)$, or equivalently with the optimal potential $V^*(x,\tau)$.

\subsection{Examples for $0\le x\le x_0$ with $x_0$ a finite number}
\noindent {\bf (I)} Let
\begin{eqnarray*}
x_0=1,\ t_h=1,\ p_0(x)=2x,\ \Gamma(x)=\int_0^xp_0(z)d\,z=x^2,
\end{eqnarray*}
one can show that
\begin{eqnarray*}
f_0(x)=\int_0^xp_0(z)dz=x^2,\quad
f_1(x)=f_0(\Gamma^{-1}(x))=(\sqrt{x})^2=x.
\end{eqnarray*}
From Eq. (\ref{eq9}) and the definition of $\Gamma(x,\tau)$, we have
\begin{eqnarray*}
&&\Gamma(x,\tau)=x+[\Gamma(x)-x]\tau=\tau x^2+ (1-\tau)x,\cr
&&\Gamma_{\tau}^{-1}(x)=2x/[\sqrt{(1-\tau)^2+4\tau x}+(1-\tau)].
\end{eqnarray*}
Therefore,
\begin{eqnarray*}
f^*(x,\tau)&=&f_0(\Gamma_{\tau}^{-1}(x))=\left(\frac{2x}{\sqrt{(1-\tau)^2+4\tau x}+(1-\tau)}\right)^2,\cr
p^*(x,\tau)&=&\frac{p_0(\Gamma_{\tau}^{-1}(x))}{\left.\partial_z\Gamma_{\tau}(z)\right|_{z=\Gamma_{\tau}^{-1}(x)}}
=\frac{2\Gamma_{\tau}^{-1}(x)}{2\tau\Gamma_{\tau}^{-1}(x)+(1-\tau)}\cr
&=&\frac{1}{\tau}+\frac{\tau-1}{\tau\sqrt{(1-\tau)^2+4\tau x}},
\end{eqnarray*}
and potential $V^*(x,\tau)$ can be obtained by Eq.~(\ref{eq29}). For this particular case, the work output in one cyclic period  (see Eq.~(\ref{eq16})) is
$$
W^*=k_B(T_h-T_c)\left(\ln2-\frac12\right)-\frac{1}{30}\left(\frac{\xi_h}{t_h}+\frac{\xi_c}{t_c}\right),
$$
and the efficiency (see Eq.~(\ref{eq24})) is
$$\eta^*=\eta_C-(1-\eta_C)\varepsilon,$$
with
$$\varepsilon=\frac{1/(t_cD_c)+1/(t_hD_h)}{30\ln2-15-1/(t_hD_h)}.$$

\noindent {\bf (II)} Next, for
$$
x_0=1,\ t_h=1,\ p_0(x)=\frac{1}{2\sqrt{x}},\
\Gamma(x)=\int_0^xp_0(z)d\,z=\sqrt{x},
$$
it can be shown that $f_0(x)=\sqrt{x}$, $f_1(x)=x$, and
\begin{eqnarray*}
\Gamma(x,\tau)=\tau \sqrt{x}+(1-\tau)x,\quad \Gamma_{\tau}^{-1}(x)=\left(\frac{\sqrt{\tau^2+4(1-\tau)x}-\tau}{2(1-\tau)}\right)^2.
\end{eqnarray*}
So
\begin{eqnarray*}
f^*(x,\tau)&=&f_0(\Gamma_{\tau}^{-1}(x))=\frac{\sqrt{\tau^2+4(1-\tau)x}-\tau}{2(1-\tau)},\cr p^*(x,\tau)&=&\frac{p_0(\Gamma_{\tau}^{-1}(x))}{\left.\partial_z\Gamma_{\tau}(z)\right|_{z=\Gamma_{\tau}^{-1}(x)}}
=\frac{1}{\sqrt{\tau^2+4(1-\tau)x}}.
\end{eqnarray*}
Similarly, $V^*(x,\tau)$ can be obtained by Eq.~(\ref{eq29}). For this particular case, the work (see Eq.~(\ref{eq16})) is $$
W^*=k_B(T_h-T_c)(1-\ln2)-\frac{\xi_h/t_h+\xi_c/t_c}{30},
$$
and the efficiency (see Eq.~(\ref{eq24})) is
$$
\eta^*=\eta_C-(1-\eta_C)\varepsilon,
$$
with
$$
\varepsilon=\frac{1/(t_cD_c)+1/(t_hD_h)}{30-30\ln2-1/(t_hD_h)}.
$$

\subsection{An example for $0\le x< \infty$}
We now give an example for $x_0=\infty$. Let $t_h=1$, and
\begin{eqnarray*}
p_0(x)=\frac{kx^{k-1}}{\lambda_0^k} \exp\left[-\left(\frac{x}{\lambda_0}\right)^k\right],\quad p_1(x)=\frac{kx^{k-1}}{\lambda_1^k} \exp\left[-\left(\frac{x}{\lambda_1}\right)^k\right],
\end{eqnarray*}
{\it i.e.,} both $p_0(x)$ and $p_1(x)$ are Weibull distributions. Then for the optimal case,  one can get $\Gamma(x)=\lambda_1x/\lambda_0$, and
\begin{eqnarray*}
\Gamma(x,\tau)=(1-\tau+\tau\lambda_1/\lambda_0)x,\quad \Gamma_{\tau}^{-1}(x)=\frac{x}{1-\tau+\tau\lambda_1/\lambda_0}.
\end{eqnarray*}
Consequently,
\begin{eqnarray*}
f^*(x,\tau)&=&f_0(\Gamma_{\tau}^{-1}(x))=1-\exp\left[-\left(\frac{x}{\lambda(\tau)}\right)^k\right],\cr
p^*(x,\tau)&=&\frac{kx^{k-1}}{[\lambda(\tau)]^k} \exp\left[-\left(\frac{x}{\lambda(\tau)}\right)^k\right],
\end{eqnarray*}
with $\lambda(\tau)=(1-\tau)\lambda_0+\tau\lambda_1$.
From Eq.~(\ref{eq22}), the work at maximum power is
$$
\begin{aligned}
W^*_{\max P}=\frac{k_B(T_h-T_c)}{2}\int_0^\infty p_0(x)\ln\Gamma'(x) d\,x
=\frac{k_B(T_h-T_c)}{2}\ln\left(\frac{\lambda_1}{\lambda_0}\right),
\end{aligned}
$$
which decreases with $\lambda_0$ while increases with $\lambda_1$.

For $\lambda_0\to0$, $p_0(x)$ tends to Dirac delta function $\delta(x)$. While for $\lambda_1\to\infty$, $p_1(x)$ tends to a constant. This is consistent with previous discussion for finite scale parameter $x_0$ cases, in which maximum of $W^*_{\max P}$ is reached when $p_1(x)\equiv1/x_0$ is constant. As before, the optimal potential $V^*(x,\tau)$ can be obtained by Eq.~(\ref{eq29}).
The work for this particular case is (see Eq.~(\ref{eq16}))
\begin{equation}\label{eq35}
W^*=k_B(T_h-T_c)\ln\left(\frac{\lambda_1}{\lambda_0}\right)-(\lambda_1-\lambda_0)^2\left(\frac{\xi_h}{t_h}+\frac{\xi_c}{t_c}\right){\bf \Gamma}\left(1+\frac{2}{k}\right),
\end{equation}
with ${\bf \Gamma}(x)$ the Gamma function defined as
$$
{\bf \Gamma}(x)=\int_0^{+\infty}t^{x-1}e^{-t}dt.
$$

For given value of parameter $\lambda_0$ in the expression of initial probability $p_0(x)$, there exists an optimal parameter
$$
\lambda_1^*=\frac{\lambda_0+\sqrt{\lambda_0^2+\frac{2k_B(T_h-T_c)}{(\xi_h/t_h+\xi_c/t_c)\mathfrak{F}(1+2/k)}}}{2},
$$
with which work $W^*$ reaches its maximum. The efficiency for these particular cases (see Eq.~(\ref{eq24}) for definition) is also $\eta^*=\eta_C-(1-\eta_C)\varepsilon$ but with
$$
\varepsilon=\frac{\left(\frac{1}{t_cD_c}+\frac{1}{t_hD_h}\right)\mathfrak{F}\left(1+\frac{2}{k}\right)}{\frac{\ln(\lambda_1/\lambda_0)}{(\lambda_1-\lambda_0)^2}-\frac{\mathfrak{F}(1+2/k)}{t_hD_h}}.
$$
One can easily show that $\varepsilon$ increases with $\lambda_1$ for $\lambda_1\ge\lambda_0$, and $\eta^*\to\eta_C$ when $\lambda_1\to\lambda_0$, but with which the work $W^*\to0$.

\subsection{An example for $-\infty\le x< \infty$}
Finally, we give an example in which state (spatial) variable $x$ is defined in the whole real domain, {\it i.e.}, $x\in(-\infty$, $+\infty$). As before, for convenience, let $t_h=1$, but assume the following initial and final probabilities
\begin{eqnarray*}
p_0(x)=\frac{1}{2\beta_0}\exp\left(-\frac{|x-\alpha_0|}{\beta_0}\right),\quad p_1(x)=\frac{1}{2\beta_1}\exp\left(-\frac{|x-\alpha_1|}{\beta_1}\right).
\end{eqnarray*}
Usually, distributions with probability like these are called Laplace distributions. For this example, $\Gamma(x)=\beta_1(x-\alpha_0)/\beta_0+\alpha_1$, and
\begin{eqnarray*}
\Gamma(x,\tau)&=&(1-\tau+\beta_1\tau/\beta_0)x+(\alpha_1-\alpha_0\beta_1/\beta_0)\tau,\cr \Gamma_{\tau}^{-1}(x)&=&\frac{\beta_0x+(\alpha_0\beta_1-\alpha_1\beta_0)\tau}{(1-\tau)\beta_0+\tau\beta_1}.
\end{eqnarray*}
So the optimal {\it distribution function}
$$
f^*(x,\tau)=\frac12+\frac12\textrm{sgn}\textbf{(}x-\alpha(\tau)\textbf{)}
\left[1-\exp\left(-\frac{|x-\alpha(\tau)|}{\beta(\tau)}\right)\right],
$$
with
$$
\alpha(\tau)=(1-\tau)\alpha_0+\tau\alpha_1,\quad \beta(\tau)=(1-\tau)\beta_0+\tau\beta_1.
$$
Here $\textrm{sgn}(x)$ is the sign function. It can be shown that the optimal probability is
$$
p^*(x,\tau)=\frac{1}{2\beta(\tau)}\exp\left(-\frac{|x-\alpha(\tau)|}{\beta(\tau)}\right).
$$

For these particular cases, the output work is
\begin{equation}\label{eq36}
W^*=k_B(T_h-T_c)\ln\left(\frac{\beta_1}{\beta_0}\right)-\left(\frac{\xi_h}{t_h}+\frac{\xi_c}{t_c}\right)[2(\beta_1-\beta_0)^2+(\alpha_1-\alpha_0)^2].
\end{equation}
Compared with the results for Weibull distribution with parameter $k=1$ ({\bf Note,} $\mathfrak{F}(3)=2$), an extra term $(\xi_h/t_h+\xi_c/t_c)(\alpha_1-\alpha_0)^2$ is added to the {\it irreversible work} $W^*_{irr}$, see Eqs.~(\ref{eq11}, \ref{eq12}, \ref{eq35}, \ref{eq36}). This is due to the spatial translocation of the system from one mean position $\int_{-\infty}^{\infty}xp_0(x)dx=\alpha_0$ to another mean position $\int_{-\infty}^{\infty}xp_1(x)dx=\alpha_1$. If $\alpha_0=\alpha_1$, then properties of work $W^*$ and efficiency $\eta^*$ are the same as the ones discussed in the above subsection for Weibull distribution with parameter $k=1$.

\section{Conclusions and Remarks}
In this study, optimization methods for stochastic thermodynamic machines are discussed, and general properties of optimal machines are analyzed. To illustrate the results, several explicit examples with optimal protocol are also presented. The main idea of optimization is to find the optimal external potential, with which the heat dissipation into environment reaches its minimum. It should be noted that, within the theoretical framework of Langevin stochastic processes and using the method of Monge-Kantorovich optimal mass transport, similar results about the minimum of heat dissipation have been obtained generally in \cite{Aurell2011,Aurell2012,Bo2013}. But differently, in this study, the minimum of heat dissipation is obtained by the method of characteristics and variation, and within the framework of Fokker-Planck equation, which makes it easier to construct illustrating examples and to know more clearly what protocol (external potential) we should choose to reach the optimal one.

It seems that several results obtained in this study are similar as the ones have been presented in previous references, including those for efficiency and work at maximum power. But with the optimal {\it distribution function} $f^*(x,\tau)$, and consequently with the optimal potential $V^*(x,\tau)$, the period $t^*=2t^{\rm st}$ required to get the maximum power is less than those given in previous references. In fact, this study indicates that the period $t^*=2t^{\rm st}$, see Eq.~(\ref{eq19-2}), required to get the maximum power is the least one. In other words, the thermodynamic machines designed in this study, with optimal potential $V^*(x,\tau)$, can output work more fast and have the smallest {\it stalling time} $t^{\rm st}$ than any other ones.
One of the main contributions of this study is that, the least value of {\it irreversible work} $W_{irr}^*$, or equivalently the least value of entropy production $\Delta S_{\rm tot}$, dissipated in one work cycle is obtained, and the optimal protocol which can lead to this lower bound is also presented.

\vspace*{1cm}

\setcounter{figure}{0}
\setcounter{section}{0}
\setcounter{equation}{0}
\setcounter{table}{0}

\renewcommand{\thefigure}{A\arabic{figure}}
\renewcommand{\thesection}{A\arabic{section}}
\renewcommand{\theequation}{A\arabic{equation}}
\renewcommand{\thetable}{A\Roman{table}}

\centerline{\bf Appendix: Details of analysis on maximum power}\label{appendex}
\vspace*{0.5cm}
For convenience, in this appendix, the following notations will be employed,
\begin{equation}\label{S1}
\begin{aligned}
B:=&k_B(T_h-T_c)\int_0^{x_0}p^*_0(x)\ln \Gamma'(x) dx,\cr
A:=&\left(\frac{\xi_h}{\bar{t}_h}+\frac{\xi_c}{\bar{t}_c}\right)\int_0^{x_0}p_0^*(x)[\Gamma(x)-x]^2,\cr
C:=&\frac{{\xi_h}/{\bar{t}_h}}{{\xi_h}/{\bar{t}_h}+{\xi_c}/{\bar{t}_c}}.
\end{aligned}
\end{equation}
With these notations, and from Eqs.~(\ref{eq15}, \ref{eq16}, \ref{eq19-2}), one can easily show that {\it stalling time} $t^{\rm st}=A/B$, {\it optimal period} $t^*=2t^{\rm st}=2A/B$, work $W^*(t)=B-A/t$, heat $Q^*_h=B/\eta_C-AC/t$, power $P(t)=W^*(t)/t=B/t-A/t^2$, and the maximum power $P^*=P(t^*)=W^*(t^*)/t^*=B^2/4A$. Therefore,
\begin{eqnarray}\label{S2}
\frac{P(t)}{P^*}=\frac{t^*}{t}\left(2-\frac{t^*}{t}\right)=1-\left(\frac{t^*}{t}-1\right)^2,
\end{eqnarray}
which equals to zero when $t=t^{\rm st}=t^*/2$ or $t\to\infty$, and reaches its maximum 1 for $t=t^*$. Meanwhile, with notations in Eq.~(\ref{S1}), the efficiency $\eta^*=W^*/Q_h^*$ can be written as $\eta^*=(B-A/t)/(B/\eta_C-AC/t)$, which gives that
\begin{eqnarray}\label{S3}
\frac{t^*}{t}=\frac{2(1-\eta^*/\eta_C)}{1-C\eta^*},\quad
\eta^*=\frac{(2-t^*/t)\eta_C}{2-C\eta_Ct^*/t}.
\end{eqnarray}
Since $C\eta_C<1$, it can be shown that efficiency $\eta^*$ increases monotonically from 0 to $\eta_C$ as the period $t$ increases from $t^{\rm st}$ to infinity. For $t=t^*$, {\it i.e.}, when power $P(t)$ reaches its maximum $P^*=B^2/4A$, efficiency $\eta^*$ get its value $\eta_{\max P}^*=\eta_C/(2-C\eta_C)$.

From Eqs.~(\ref{S2}, \ref{S3}), the ratio ${P(t)}/{P^*}$ can be rewritten as follows
\begin{equation}\label{S4}
\begin{aligned}
\frac{P(t)}{P^*}=&\frac{\eta(1-\eta/\eta_C)}{(1-C\eta)^2}\cdot\frac{4(1-C\eta_C)}{\eta_C}\cr
=&4\left(\frac{1}{\eta_C}-\frac{{\xi_h}/{\bar{t}_h}}{{\xi_h}/{\bar{t}_h}+{\xi_c}/{\bar{t}_c}}\right)
\cdot\frac{\eta(1-\eta/\eta_C)}{\left(1-\frac{{\xi_h}/{\bar{t}_h}}{{\xi_h}/{\bar{t}_h}+{\xi_c}/{\bar{t}_c}}\eta\right)^2},
\end{aligned}
\end{equation}
which equals to zero if $\eta=0$ or $\eta=\eta_C$, and reaches its maximum 1 if $\eta=\eta_C/(2-C\eta_C)$.

From Eqs.~(\ref{eq22}, \ref{S1}), the work at maximal power $P^*$ is $W^*_{\max P}=B/2$. So $W^*(t)/W^*_{\max P}=2-t^*/t=2(1-C\eta_C)\eta^*/[(1-C\eta^*)\eta_C]$. Obviously, $W^*(t)/W^*_{\max P}$ increases monotonically from 0 to 2 as period $t$ increases from {\it stalling time } $t^{\rm st}=t^*/2$ to infinity, or equivalently as efficiency $\eta^*$ increases from 0 to Carnot efficiency $\eta_C$.


\begin{thebibliography}{10}

\bibitem{Carnot1824}
S.~Carnot.
\newblock {\em Refl\'{e}xions sur la Puissance Motrice du Feu, etsur les
  Machines Propres \`{a} D\'{e}velopper cette Puissance}.
\newblock 1824.

\bibitem{Clausius1856}
R.~Clausius.
\newblock On a modified form of the second fundamental theorem in the
  mechanical theory of heat.
\newblock {\em Phil. Mag. Ser.}, 4(12):81--98, 1856.

\bibitem{Yvon1955}
J.~Yvon.
\newblock {\em Proceedings of the International Conference on Peaceful Uses of
  Atomic Energy}.
\newblock United Nations, Geneva, 1955, p.~387.

\bibitem{Chambadal1957}
P.~Chambadal.
\newblock {Les Centrales Nucl\'{e}aries}.
\newblock {\em Armand Colin}, 11:41--58, 1957.

\bibitem{Novikov1958}
I.~I. Novikov.
\newblock The efficiency of atomic power stations (a review).
\newblock {\em J. Nucl. Energy}, 7:125--128, 1958.

\bibitem{Curzon1975}
F.~L. Curzon and B.~Ahlborn.
\newblock Effciency of a carnot engine at maximum power output.
\newblock {\em Phil. Mag. Ser.}, 43:22--24, 1975.

\bibitem{Broeck2005}
C.~Van den Broeck.
\newblock Thermodynamic efficiency at maximum power.
\newblock {\em Phys. Rev. Lett.}, 95(19):190602, 2005.

\bibitem{Schmiedl2008Efficiency1}
T.~Schmiedl and U.~Seifert.
\newblock Efficiency at maximum power: An analytically solvable model for
  stochastic heat engines.
\newblock {\em Europhys. Lett.}, 81(2):20003, 2008.

\bibitem{Esposito2010Efficiency}
M.~Esposito, R.~Kawai, K.~Lindenberg, and C.~Van den Broeck.
\newblock Efficiency at maximum power of low-dissipation carnot engines.
\newblock {\em Phys. Rev. Lett.}, 105(15):150603, 2010.

\bibitem{Izumida2011Efficiency}
Y.~Izumida and K.~Okuda.
\newblock Efficiency at maximum power of minimally nonlinear irreversible heat
  engines.
\newblock {\em Europhys. Lett.}, 97(1):10004, 2011.

\bibitem{Proesmans2016}
K.~Proesmans, B.~Cleuren, and C.~Van den Broeck.
\newblock Power-efficiency-dissipation relations in linear thermodynamics.
\newblock {\em Phys. Rev. Lett.}, 116:220601, 2016.

\bibitem{Holubec2016}
V.~Holubec and A.~Ryabov.
\newblock Maximum efficiency of low-dissipation heat engines at arbitrary
  power.
\newblock {\em J. Stat. Mech.}, 2016(7):073204, 2016.

\bibitem{Ryabov2016}
A.~Ryabov and V.~Holubec.
\newblock Maximum efficiency of steady-state heat engines at arbitrary power.
\newblock {\em Phys. Rev. E}, 93(5):050101, 2016.

\bibitem{Pietzonka2018}
P.~Pietzonka and U.~Seifert.
\newblock Universal trade-off between power, efficiency, and constancy in
  steady-state heat engines.
\newblock {\em Phys. Rev. Lett.}, 120:190602, 2018.

\bibitem{Benenti2011}
G.~Benenti, K.~Saito, and G.~Casati.
\newblock Thermodynamic bounds on efficiency for systems with broken
  time-reversal symmetry.
\newblock {\em Phys. Rev. Lett.}, 106:230602, 2011.

\bibitem{Golubeva2012Efficiency}
N.~Golubeva and A.~Imparato.
\newblock Efficiency at maximum power of interacting molecular machines.
\newblock {\em Phys. Rev. Lett.}, 109(19):190602, 2012.

\bibitem{Allahverdyan2013}
A.~E. Allahverdyan, K.~V. Hovhannisyan, A.~V. Melkikh, and S.~G. Gevorkian.
\newblock Carnot cycle at finite power: Attainability of maximal efficiency.
\newblock {\em Phys. Rev. Lett.}, 111:050601, 2013.

\bibitem{Holubec2015}
V.~Holubec and A.~Ryabov.
\newblock Efficiency at and near maximum power of low-dissipation heat engines.
\newblock {\em Phys. Rev. E}, 92:052125, 2015.

\bibitem{Calvo2015}
A.~Calvo Hern\'{a}ndez, A.~Medina, and J.~M.~M. Roco.
\newblock Time, entropy generation, and optimization in low-dissipation heat
  devices.
\newblock {\em New J. Phys.}, 17(1):075011, 2015.

\bibitem{Reyes2017}
I.~Reyes-Ram\'{\i}rez, J.~Gonzalez-Ayala, A.~Calvo Hern\'andez, and
  M.~Santill\'an.
\newblock Local-stability analysis of a low-dissipation heat engine working at
  maximum power output.
\newblock {\em Phys. Rev. E}, 96:042128, 2017.

\bibitem{Polettini2017}
M.~Polettini and M.~Esposito.
\newblock Carnot efficiency at divergent power output.
\newblock {\em Europhys. Lett.}, 118(4):40003, 2017.

\bibitem{Sekimoto2010}
K.~Sekimoto.
\newblock {\em Stochastic Energetics}.
\newblock Springer Berlin Heidelberg, 2010.

\bibitem{Seifert2012Stochastic}
U.~Seifert.
\newblock Stochastic thermodynamics, fluctuation theorems and molecular
  machines.
\newblock {\em Rep. Prog. Phys.}, 75(75):126001, 2012.

\bibitem{Mart2016Colloidal}
I.~A. Mart\'{\i}nez, \'{E}. Rold\'{a}n, L.~Dinis, and R.~A. Rica.
\newblock Colloidal heat engines: a review.
\newblock {\em Soft Matter}, 13(1):22, 2016.

\bibitem{Giuliano2017}
B.~Giuliano, G.~Casati, K.~Saito, and R.~Whitney.
\newblock Fundamental aspects of steady-state conversion of heat to work at the
  nanoscale.
\newblock {\em Phys. Rep.}, 694:1--124, 2017.

\bibitem{Mart2016Brownian}
I.~A. Mart\'{\i}nez, Rold\'{a}n, L~Dinis, D~Petrov, J.~M. Parrondo, and R.~A.
  Rica.
\newblock Brownian carnot engine.
\newblock {\em Nat. Phys.}, 12(1):67--70, 2016.

\bibitem{Lee2017Carnot}
J.~S. Lee and H.~Park.
\newblock Carnot efficiency is reachable in an irreversible process.
\newblock {\em Sci. Rep.}, 7(1):10725, 2017.

\bibitem{Holubec2018}
V.~Holubec and A.~Ryabov.
\newblock Cycling tames power fluctuations near optimum efficiency.
\newblock {\em Phys. Rev. Lett.}, 121:120601, 2018.

\bibitem{Hondou2000Unattainability}
T.~Hondou and K.~Sekimoto.
\newblock Unattainability of carnot efficiency in the brownian heat engine.
\newblock {\em Phys. Rev. E}, 62:6021--6025, 2000.

\bibitem{Verley2014The}
G.~Verley, M.~Esposito, T.~Willaert, and C.~Van den Broeck.
\newblock The unlikely carnot efficiency.
\newblock {\em Nat. Commun.}, 5:4721, 2014.

\bibitem{Schmiedl2007}
T.~Schmiedl and U.~Seifert.
\newblock Optimal finite-time processes in stochastic thermodynamics.
\newblock {\em Phys. Rev. Lett.}, 98:108301, 2007.

\bibitem{Holubec2014}
V.~Holubec.
\newblock An exactly solvable model of a stochastic heat engine: optimization
  of power, power fluctuations and efficiency.
\newblock {\em J. Stat. Mech.}, 5(5), 2014.

\bibitem{Zhang2019}
Y.~Zhang.
\newblock Optimisation of a class of heat engines with explicit solution.
\newblock {\em Physica A}, 527:121272, 2019.

\bibitem{Then2008}
H.~Then and A.~Engel.
\newblock Computing the optimal protocol for finite-time processes in
  stochastic thermodynamics.
\newblock {\em Phys. Rev. E}, 77:041105, 2008.

\bibitem{Horowitz2018}
J.~M. Horowitz and A.~P. Solon.
\newblock Phase transition in protocols minimizing work fluctuations.
\newblock {\em Phys. Rev. Lett.}, 120:180605, 2018.

\bibitem{Van2005Thermodynamic}
C.~Van den Broeck.
\newblock Thermodynamic efficiency at maximum power.
\newblock {\em Phys. Rev. Lett.}, 95(19):190602, 2005.

\bibitem{Cleuren2009Universality}
B.~Cleuren, B.~Rutten, and C.~Van den Broeck.
\newblock Universality of efficiency at maximum power.
\newblock {\em Phys. Rev. Lett.}, 102(13):879--889, 2009.

\bibitem{Van2012Efficiency}
C.~Van den Broeck, N.~Kumar, and K.~Lindenberg.
\newblock Efficiency of isothermal molecular machines at maximum power.
\newblock {\em Phys. Rev. Lett.}, 108(21):210602, 2012.

\bibitem{Wang2013Efficiency}
R.~Wang, J.~Wang, J.~He, and Y.~Ma.
\newblock Efficiency at maximum power of a heat engine working with a two-level
  atomic system.
\newblock {\em Phys. Rev. E}, 87(4):042119, 2013.

\bibitem{Hooyberghs2013Efficiency}
H.~Hooyberghs, B.~Cleuren, A.~Salazar, J.~O. Indekeu, and C.~Van den Broeck.
\newblock Efficiency at maximum power of a chemical engine.
\newblock {\em J. Chem. Phys.}, 139(13):22, 2013.

\bibitem{Loeb2016}
P.~A. Loeb.
\newblock {\em Real Analysis}.
\newblock Birkhauser Verlag AG, 2016.

\bibitem{Brandner2015}
K.~Brandner, K.~Saito, and U.~Seifert.
\newblock Thermodynamics of micro- and nano-systems driven by periodic
  temperature variations.
\newblock {\em Phys. Rev. X}, 5:031019, 2015.

\bibitem{Proesmans2015}
K.~Proesmans and C.~Van~den Broeck.
\newblock Onsager coefficients in periodically driven systems.
\newblock {\em Phys. Rev. Lett.}, 115:090601, 2015.

\bibitem{Gonzalezayala2016}
J.~Gonzalezayala, A.~C. Hern\'{a}ndez, and J.~M.~M. Roco.
\newblock Irreversible and endoreversible behaviors of the ld-model for heat
  devices: the role of the time constraints and symmetries on the performance
  at maximum¦Öfigure of merit.
\newblock {\em J. Stat. Mech.}, 2016:073202, 2016.

\bibitem{Johal2017Heat}
R.~S. Johal.
\newblock Heat engines at optimal power: Low-dissipation versus endoreversible
  model.
\newblock {\em Phys. Rev. E}, 96:012151, 2017.

\bibitem{Aurell2011}
E.~Aurell, C.~Mej\'{\i}a-Monasterio, and P.~Muratore-Ginanneschi.
\newblock Optimal protocols and optimal transport in stochastic thermodynamics.
\newblock {\em Phys. Rev. Lett.}, 106:250601, 2011.

\bibitem{Aurell2012}
E.~Aurell, K.~Gaw\c{e}dzki, C.~Mej\'{\i}a-Monasterio, R.~Mohayaee, and
  P.~Muratore-Ginanneschi.
\newblock Refined second law of thermodynamics for fast random processes.
\newblock {\em J. Stat. Phys.}, 147:487--505, 2012.

\bibitem{Bo2013}
S.~Bo, E.~Aurell, R.~Eichhorn, and A.~Celani.
\newblock Optimal stochastic transport in inhomogeneous thermal environments.
\newblock {\em Europhys. Lett.}, 103:10010--10015, 2013.

\end{thebibliography}
\end{document}